\documentclass[reprint, superscriptaddress]{revtex4-2}

\usepackage[toc,page]{appendix}
\usepackage[british]{babel}
\usepackage[utf8]{inputenc}
\usepackage{amsmath,amssymb}
\usepackage{graphicx}
\usepackage{ae}
\usepackage{esdiff}
\usepackage{braket}
\usepackage{bbm}
\usepackage{simplewick}
\usepackage{float}
\usepackage{braket}
\usepackage{simplewick}
\usepackage{color}
\usepackage{framed}
\usepackage[caption=false]{subfig}
\usepackage[pdftex,colorlinks=true,linkcolor=blue,citecolor=blue,urlcolor=black]{hyperref}
\usepackage{mathtools}
\usepackage{enumitem}
\usepackage{ragged2e}
\usepackage{floatflt}
\usepackage{import}
\usepackage{titlesec}
\usepackage{multirow}
\usepackage{etoolbox}
\usepackage{appendix}
\usepackage{dsfont}
\usepackage{comment}
\usepackage{xcolor}
\usepackage{blindtext}
\usepackage{soul,xcolor}

\begin{document}

\newcommand{\norm}[1]{\left\lVert#1\right\rVert}

\setstcolor{red}

\def \r{{\boldsymbol{r}}}
\def \k{{\boldsymbol{k}}}
\def \p{{\boldsymbol{p}}}
\def \q{{\boldsymbol{q}}}
\def \x{{\textbf{x}}}
\def \A{{\textbf{{A}}}}
\def \a{{\textbf{{a}}}}
\def \b{{\textbf{{b}}}}
\def \z{{\textbf{{z}}}}
\def \0{{\boldsymbol{{0}}}}
\def \dl{\frac{\partial}{\partial l}}
\def \P{{\boldsymbol{P}}}
\def \K{{\boldsymbol{K}}}
\def \sigmad{{\sigma_{\downarrow\uparrow}}}
\def \uone{{\boldsymbol{u}_1}}
\def \utwo{{\boldsymbol{u}_2}}
\def \edown{{\epsilon_{\downarrow}}}

\def \intk{{\int_\textbf{k}}}
\def \eone{{\epsilon_{\uparrow,1}}}
\def \etwo{{\epsilon_{\uparrow,2}}}
\def \Eone{{\epsilon_{\downarrow,1}}}

\definecolor{mgrey}{RGB}{63,63,63}
\definecolor{mred}{RGB}{235,97,51}
\newcommand{\mg}[1]{{\color{mgrey}{#1}}}
\newcommand{\mr}[1]{{\color{mred}{#1}}}
\newcommand{\red}[1]{{\color{red}{#1}}}

\def\BigColSep{\setlength{\arraycolsep}{50pt}}

\title{Topological transport of mobile impurities}

\author{D.\ Pimenov}\email{dpimenov@umn.edu}
\affiliation{William I. Fine Theoretical Physics Institute, University of Minnesota, Minneapolis, MN 55455, USA}
\author{A.\ Camacho-Guardian} 
\affiliation{Department of Physics and Astronomy, Aarhus University, Ny Munkegade, DK-8000 Aarhus C, Denmark}
\affiliation{T.C.M. Group, Cavendish Laboratory, University of Cambridge,
JJ Thomson Avenue, Cambridge, CB3 0HE, U.K.}
\author{N.\ Goldman} 
\affiliation{Universit\'e Libre de Bruxelles, CP 231, Campus Plaine, 1050 Brussels, Belgium}
\author{P. Massignan}
\affiliation{Department de F\'isica, Universitat Polit\`ecnica de Catalunya, Campus Nord, B4-B5, E-08034 Barcelona, Spain}
\author{G.\ M.\ Bruun}
\affiliation{Department of Physics and Astronomy, Aarhus University, Ny Munkegade, DK-8000 Aarhus C, Denmark}
\author{M.\ Goldstein}
\affiliation{Raymond and Beverly Sackler School of Physics and Astronomy, Tel Aviv University, Tel Aviv 6997801, Israel}

\begin{abstract}
We study the Hall response of topologically-trivial mobile impurities (Fermi polarons) interacting weakly with majority fermions forming a Chern-insulator background. This setting involves a rich interplay between the genuine many-body character of the polaron problem and the topological nature of the surrounding cloud. When the majority fermions are accelerated by an external field, a transverse impurity current can be induced. To quantify this polaronic Hall effect, we compute the drag transconductivity, employing controlled diagrammatic perturbation theory in the impurity-fermion interaction. We show that the impurity Hall drag is not simply proportional to the Chern number characterizing the topological transport of the insulator on its own -- it also depends continuously on particle-hole breaking terms, to which the Chern number is insensitive. However, when the insulator is tuned across a topological phase transition, a sharp jump of the impurity Hall drag results, for which we derive an analytical expression. We describe how to experimentally detect the polaronic Hall drag and its characteristic jump, setting the emphasis on the circular dichroism displayed by the impurity's absorption rate.
\end{abstract}

 \maketitle

\section{Introduction}
\label{Introsec}

As a rule of thumb, interacting many-body systems in more than one dimension are difficult to analyze, and controllable routes  to the inclusion of interactions are rare. One such approach is to consider a non-interacting ``majority" system, couple it to a small number of quantum impurities, and study interaction effects on the impurities only. If the majority system is a conventional metal, the impurities are transformed into so-called Fermi polarons \cite{chevy2006universal} \footnote{Here, we understand the polaron as a mobile quasiparticle, and not as a static impurity as recently studied in a topological system in Ref.\ \cite{PhysRevLett.125.240601}}, which by now are routinely observed in ultracold-gas \cite{schirotzek2009observation,Kostall2012,Koschorreck2012,PhysRevLett.118.083602,PhysRevLett.122.093401} and also solid state experiments \cite{sidler2017fermi} -- for a review, see for instance Refs.\ \cite{massignan2014polarons, Levinsen2015, schmidt2018universal}. 

In these systems, the local kinematic properties of the impurities are modified by the interaction with the medium, while the medium itself is unmodified if the impurity density is small. The next logical question to ask is whether global topological characteristics of the medium \cite{RevModPhys.89.040502} can influence the impurity as well: Can a topologically trivial impurity inherit the topological quantum numbers of the medium? Such an interaction-induced topology is a fundamentally interesting  prospect. Furthermore, this question is of high relevance to current cold-atom experiments, where a broad family of topological band structures have been realized \cite{RevModPhys.91.015005}. Topological and polaronic physics are thus well-controlled (and highly active) but largely separate fields in cold-atom research, and it is thus worthwhile and intriguing to combine them together. 
This goal has been approached in a few recent theoretical works, mainly from two perspectives: Either interaction effects are strong such that an impurity-majority bound state is formed \cite{Grusdt2016, PhysRevB.100.075126, PhysRevX.10.041058, baldelli2021tracing}, and the impurity inherits the topological quantum numbers of the majority, or, alternatively, one can study the problem in weak coupling \cite{PhysRevB.99.081105, *PhysRevB.102.119903}, as previously done by some of us, with the majority forming a Chern insulator. This perturbative approach is well-controlled and does not require additional regularization.

As a diagnostic tool for the inherited topological properties of the impurity particles, Ref.\ \cite{PhysRevB.99.081105, *PhysRevB.102.119903} numerically computed the impurity Hall drag for majority particles governed by the Haldane lattice model \cite{PhysRevLett.61.2015}. It was found that the Hall drag is neither quantized nor simply follows the majority phase diagram, and even vanishes in the center of the topological phase; however, it exhibits a sharp jump upon tuning the insulator across its topological phase transition. In this work, we introduce a generic (continuum) Dirac model of a Chern insulator. This model follows the same universal physics as the Haldane model, but allows for an analytical understanding of the phenomena numerically observed in Ref.\ \cite{PhysRevB.99.081105, *PhysRevB.102.119903}. 

With a diagrammatical approach, we show that the Hall drag can be split into two drag contributions exerted by majority particles and holes, respectively. These two contributions counteract each other, and completely cancel at the particle-hole symmetric point. This is reminiscent of Coulomb drag in two-layer systems \cite{kamenev1995coulomb, PhysRevB.76.081401,RevModPhys.88.025003}, and explains the observed vanishing of the drag in the center of the majority topological phase. If particle-hole symmetry is broken, the impurity Hall drag can be non-vanishing even if the majority Chern insulator is in the trivial phase. To understand the observed jump across the topological phase transition, one should view the majority system as a combination of Dirac-like fermions with linear dispersion, and ``spectator'' fermions \cite{bernevig2013topological} with a quadratic dispersion. At the phase transition, the spectator fermions change smoothly, but the Dirac fermions feel the gap closing and exhibit a singular Berry curvature. We show that this singularity is integrated over in the expression for the impurity Hall drag, which leads to a jump proportional to the change in Chern number, including the correct sign. This is the only clear manifestation of topology in weak-coupling impurity transport. We derive an analytical formula for the jump, and validate all results numerically for the Haldane lattice model. 

To supplement the theoretical results, we present a detailed discussion on how to detect the Hall drag and jump with various experimental techniques. A particular promising approach is to use circular dichroism, that is, measuring impurity excitation rates upon driving the system with left and right circularly polarized fields \cite{tran2017probing, PhysRevA.97.061602, PhysRevLett.122.166801, asteria2019measuring}. A systematic method of computing the excitation rates in an interacting many-body system is presented along the way.

The remainder of this paper is structured as follows: In Sec.\ \ref{dragDiracsec} we present the continuum Dirac model and the evaluation of the impurity drag. In Sec.\ \ref{jumpsec}, we investigate the jump across the topological phase transition. The drag including its jump at the topological transition is analyzed for the Haldane model in Sec.\ \ref{Haldanemodel}. The different measurement protocols are detailed in Sec.~\ref{dichroismsec}, with special focus on the dichroic measurement. Conclusions and outlook are presented in Sec.\ \ref{conclusionsec}. Some technical details are relegated to Appendices.

\section{Drag transconductivity in the continuum model}
\label{dragDiracsec}
We start by computing the impurity drag in a generic continuum model and consider the following two-dimensional Bloch Hamiltonian for majority particles indexed by a pseudospin $\uparrow$: 
\begin{align}
\label{contham}
&H_\uparrow(\k)  = \sum_{i = 0}^3 \psi_\uparrow^\dagger(\k) h_i(\k)  \sigma_i  \psi_\uparrow(\k)\ , \\  \notag &\psi_\uparrow(\k) = \left(c_{\uparrow,A}(\k), c_{\uparrow,B} (\k)\right)^T\ ,  \\ \notag 
& h_1(\k) = k_x\ , \quad h_2(\k) = k_y\ , \quad h_3(\k) = m + d_1k^2, \\ \notag & h_0(\k) = d_2k^2, \quad k = |\k|\ , 
\end{align}
with $\sigma_0 = \mathbbm{1}$ and $\sigma_i$ with $i = 1,2,3$ being the Pauli matrices. Throughout this paper we will
 work in units where $\hbar = c = e = 1$; all quantities are measured in appropriate powers of the (inverse) physical fermion mass, while momenta are rescaled by the band velocity.  Equation \eqref{contham} can be seen as a low-energy approximation to a microscopic tight-binding Hamiltonian with a two-sublattice structure ($A,B$) and broken time-reversal invariance. The eigenenergies corresponding to \eqref{contham} read
\begin{align}
\epsilon_{\uparrow; 1,2}(\k) = h_0(\k) \mp h(\k), \quad h(\k) = \sqrt{k^2 + h_3(k)^2} \ . 
\end{align}
Without the terms $d_1, d_2$ (which have physical dimensions (mass)$^{-1}$), Eq.\  \eqref{contham} describes a gapped Dirac cone with mass gap $m$. The term $d_1$ serves as a UV regularizer and makes the dispersion quadratic at higher energies while preserving particle-hole symmetry, $\epsilon_{\uparrow,1}(\k) = - \epsilon_{\uparrow,2}(\k)$. The symmetry is broken for finite $d_2$.  We assume $|d_1| > |d_2|$, thus the lower (upper) band is filled (empty).

\begin{figure}
\centering
\includegraphics[width=.9\columnwidth]{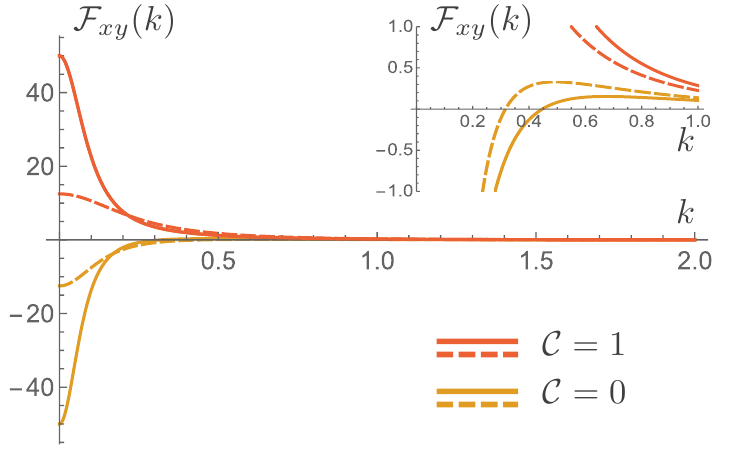}
\caption{Berry curvature for $d_1 = -1$ and $m = \pm 0.1$ (full lines),  $m = \pm 0.2$ (dashed lines). The inset shows a zoom-in on small values of $\mathcal{F}_{xy}(k)$, highlighting the sign-change of the Berry curvature in the trivial phase.}
\label{contbands}
\end{figure}

For general $d_2$, the Hamiltonian \eqref{contham} is in the Altland-Zirnbauer class A \cite{ryu2010topological}, and gives rise to a quantized Chern number $\mathcal{C}$. As shown below, it reads 
\begin{align}
\label{Cexp}
\mathcal{C}&=  \frac{1}{2\pi} \int \!d\k \frac{1}{2}\frac{(m-d_1 k^2)}{(k^2 + (m+ d_1k^2)^2)^{3/2}} \\ &= \frac{1}{2}\left[\text{sign}(m) - \text{sign}(d_1)\right]  \ . \notag
\end{align}
The integrand of Eq.\ \eqref{Cexp} is nothing but the Berry curvature $\mathcal{F}_{xy}(k)$. As visualized in Fig.\ \ref{contbands}, for $m\rightarrow 0$ $\mathcal{F}_{xy}(k)$ consists of a sharp half-quantized peak for $k\lesssim m$, arising from the Dirac fermions, on top of a broad background from high-energy  ``spectator'' fermions \cite{bernevig2013topological}. Both types of fermions effectively contribute a half-integer Chern number, such that the total Chern number is quantized to an integer.

As explicit in Eq.\ \eqref{Cexp}, $\mathcal{C}$ does not depend on the particle-hole symmetry breaking parameter $d_2$. This is in line with the geometrical interpretation of $\mathcal{C}$ as a winding number \footnote{For the winding number construction one should view momentum space as compactified, $\mathbb{R}_2 \rightarrow S_2$}, which is independent of the term $h_0$ commuting with the Hamiltonian \cite{PhysRevLett.121.086810}.

As a preparation for the later calculations, it is useful to recap the computation of $\mathcal{C}$ explicitly as  $\mathcal{C} = -2\pi {\sigma}_{xy}$ \cite{PhysRevLett.49.405, kohmoto1985topological}, with $\sigma_{xy}$ the transconductivity; the conductivity quantum is $\sigma_0 = e^2/\hbar = 1/2\pi$ with the chosen units.  In linear response, $\sigma_{xy}$ is proportional to the retarded current-current correlation function, which may be obtained by analytical continuation from imaginary time: 
\begin{align}
\label{sigmaxy1}
\sigma_{xy} = \lim_{\omega \rightarrow 0} \frac{1}{-i\omega A_0} \left[ - \braket{\hat{J}_\uparrow^x \hat{J}_\uparrow^y}(i\Omega) \bigg|_{i\Omega \rightarrow \omega + i0^+} \right], 
\end{align}
with $A_0$ the system area, and $\hat{J}_\uparrow$ the current operators at vanishing external momentum.

\begin{figure}
\centering
\includegraphics[width=.9\columnwidth]{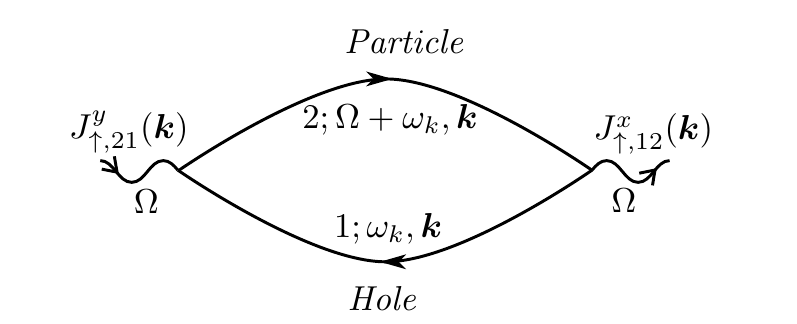}
\caption{Diagram representing Eq.\ \eqref{JJG}, with $\alpha = 1, \beta = 2$.} 
\label{bubblediag}
\end{figure}

The imaginary time correlator in Eq.\ \eqref{sigmaxy1} can be written as 
\begin{align}
\label{JJG}
&-\braket{\hat{J}_\uparrow^x \hat{J}_\uparrow^y}(i\Omega)  =  \\ \notag& A_0\int_k G_{\uparrow,\alpha}(\omega_k,\k) G_{\uparrow,\beta}(\Omega + \omega_k, \k) J^x_{\uparrow,\alpha\beta}(\k) J^y_{\uparrow,\beta\alpha}(\k),  \\ 
& \int_k \equiv \int \frac{d\k d\omega_k}{(2\pi)^3},  \quad G_{\uparrow,\alpha}(\omega_k,\k) = \frac{1}{i\omega_k - \epsilon_{\uparrow,\alpha}(\k)} \ ,  \notag 
\end{align}
where $\alpha,\beta$ refer to band indices and the Einstein summation convention is implied. $J_{\uparrow,\alpha\beta}^{x/y}$ are current matrix element in the diagonal basis (see App.\ \ref{BasisrotSec} for details). The standard diagrammatical representation of Eq.\ \eqref{JJG} is shown in Fig.\ \ref{bubblediag}. 
The Matsubara Green function $G_{\uparrow,1}$ describes the propagation of a hole in the filled lower band, while $G_{\uparrow,2}$ represents a particle in the upper band. The frequency integral in Eq.\ \eqref{JJG} only receives contributions when $\alpha \neq \beta$, and thus one can view creation of virtual particle-hole pairs as the origin of the conductivity. These quasiparticles are virtual, since the external field does not provide enough energy ($\Omega \rightarrow 0$) to overcome the band gap. 

Evaluation of Eqs.\ \eqref{JJG} and \eqref{sigmaxy1} is straightforward. One finds
\begin{align} \notag \sigma_{xy} &= -i\int\!\frac{d\k}{(2\pi)^2} \frac{J_{\uparrow,12}^x(\k) J^y_{\uparrow,21}(\k)-J_{\uparrow,21}^x(\k) J^y_{\uparrow,12}(\k)}{(\epsilon_{\uparrow,1}(\k) - \epsilon_{\uparrow,2}(\k))^2} \\&  = -\frac{1}{2\pi}  \mathcal{C} \ . 
\label{sigmaCrel}
\end{align}
Inserting current matrix elements and dispersions into Eq.\ \eqref{sigmaCrel} produces Eq.\ \eqref{Cexp}. 
After this noninteracting prelude, we are ready to attack the polaron problem. We consider a minority particle species indexed by $\downarrow$, with a trivial quadratic Hamiltonian $H_\downarrow(\p)$: 
\begin{align}
\label{trivialHam}
&H_\downarrow(\p) = \epsilon_\downarrow(\p) c^\dagger_\downarrow(\p)c_\downarrow(\p), \quad \epsilon_\downarrow(\p) = \frac{p^2}{2M} \ . 
\end{align}
We can view the impurities as governed by a similar tight-binding Hamiltonian as the majority, but with a chemical potential almost at the bottom of the lower band, around which the dispersion is approximated by an effective mass $M$. Higher impurity bands can be safely neglected. 

The majority and minority particles interact via an onsite-interaction $H_\text{int}$ \cite{PhysRevB.99.081105, *PhysRevB.102.119903}, which does not distinguish between the sublattices (recall that the sublattices give rise to the two-band structure):
 \begin{align}
&H_\text{int} = \frac{g}{A_0}\sum_{\ell = A,B}\sum_{\k,\p,\q}  c^\dagger_{\uparrow, \ell} (\k +\q) c_{\uparrow,\ell} (\k) c^\dagger_\downarrow(\p-\q) c_\downarrow(\p) =  \notag \\ \notag &
 \frac{g}{A_0} \sum_{\k,\p,\q} c^\dagger_{\uparrow, \alpha} (\k +\q) c_{\uparrow,\beta} (\k) c^\dagger_\downarrow(\p-\q) c_\downarrow(\p) W_{\alpha\beta}(\k,\q), 
 \\ &W_{\alpha\beta}(\k,\q) \equiv \left[U_\uparrow^\dagger(\k+\q) U_\uparrow(\k) \right]_{\alpha\beta}\  ,  \label{Hint}
\end{align}
where we have rotated to the band space in the second line. Now we imagine a constant and uniform force $\boldsymbol{E} = E  \textbf{e}_y$ acting on both majority and minority particles \footnote{Note that $e = 1$ is the effective charge corresponding to this force and might not be directly related to the electron charge}. Due to the interaction $H_\text{int}$, a transverse impurity current $J_{\downarrow}^x$ will be induced; without interaction, there is none due to time reversal symmetry of the impurities. To quantify this effect, we must compute the Hall drag transconductivity 
\begin{align}
\label{dragtrans1}
\sigma_{\downarrow\uparrow} \equiv \lim_{\omega \rightarrow 0} \frac{1}{-i\omega A_0} \left[ - \braket{\hat{J}_\downarrow^x \hat{J}_\uparrow^y}(i\Omega) \bigg|_{i\Omega \rightarrow \omega + i0^+} \right] \ .
\end{align}
This computation will be done to second order in the impurity-majority coupling $g$, since the first order contribution vanishes \cite{PhysRevB.99.081105, *PhysRevB.102.119903}; thus, attractive and repulsive interactions lead to the same result. We point out  that such perturbative expansion is well-controlled for small $g$, and no resummation is needed, in contrast with the recent evaluation of longitudinal polaron drag in the metallic case \cite{PhysRevX.9.041019}.

\begin{figure}[H]
\centering
\includegraphics[width=\columnwidth]{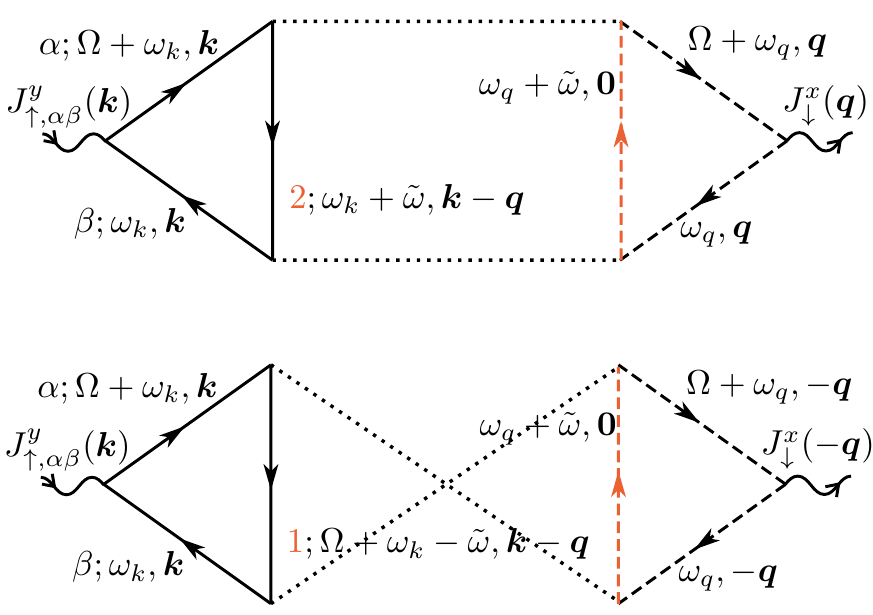}
\caption{Leading contributions to the drag transconductivity. Dashed lines represent impurities, dotted lines interaction matrix elements $W$, see Eq.\ \eqref{Hint}. The energy-momentum structure of the central part and the colored elements are explained in the main text.} 
\label{dragdiags}
\end{figure}

As in the case of Coulomb drag in two-layer systems \cite{kamenev1995coulomb}, the $\mathcal{O}(g^2)$ contribution corresponds to the two diagrams shown in Fig.\ \ref{dragdiags}. We evaluate these diagrams to leading order in the small impurity density $n_\downarrow$. The diagrams involve an impurity loop and are therefore proportional to $n_\downarrow$, unlike the single-particle polaron diagrams which have an impurity ``backbone'' \cite{PhysRevB.77.125101, *PhysRevB.77.020408}. It is convenient to identify the impurity lines that represent filled states ($\hat{=}$ impurity holes). Since these carry vanishing momenta in the small density limit, impurity lines coupled to the current vertex, $J_\downarrow^x(\q) = q_x/M$, are excluded. Thus, the central (red) line corresponds to a filled state. We may set its momentum to zero as done in Fig.\ \ref{dragdiags}, and the integration over filled states then simply produces a factor of $n_\downarrow$.

\begin{figure}[H]
\centering
\includegraphics[width=.9\columnwidth]{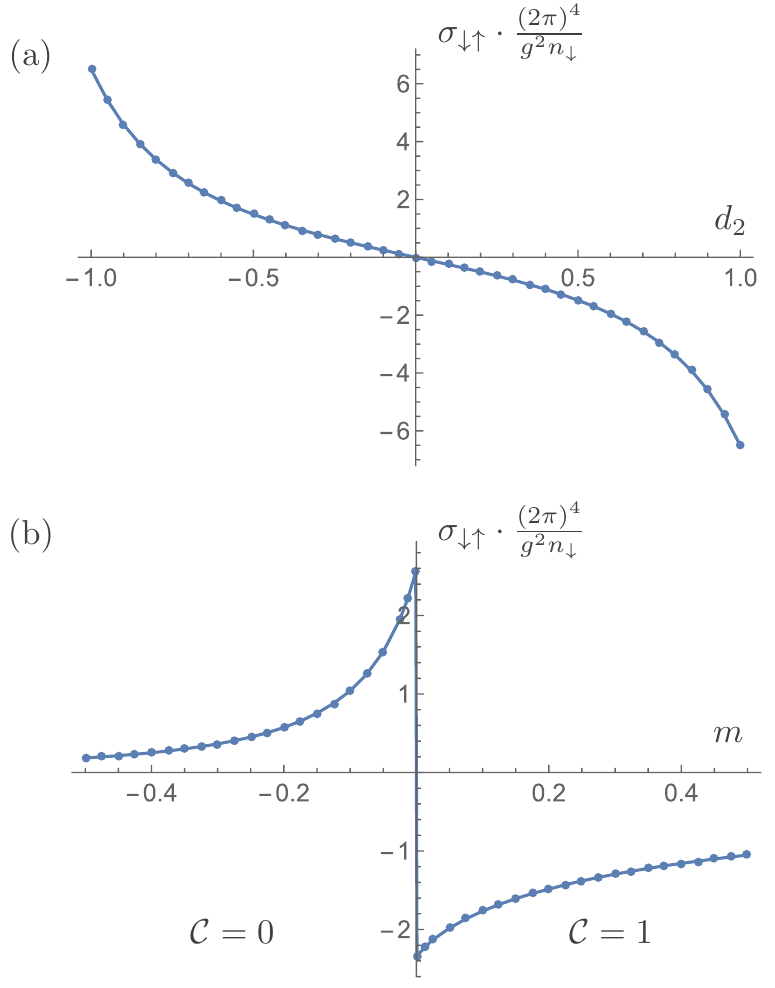}
\caption{Impurity transconductivity $\sigma_{\downarrow\uparrow}$ from numerical evaluation of Eq.\  \eqref{maineqdragcont}. Lines are guides for the eye. (a) $\sigma_{\downarrow\uparrow}$ as function of $d_2$ for $M=1, m = 0.2, d_1 = -1$. (b) $\sigma_{\downarrow\uparrow}$ as function of $m$ for $M=1, d_1 = -1, d_2 = 0.5$.}
\label{contplotnums}
\end{figure}

Identification of the red line with a filled state also fixes the (red) index of the central majority line in order for
 the $\tilde{\omega}$ integral (see Fig.\ \ref{dragdiags}) to be non-vanishing. Schematically the top diagram in Fig.\ \ref{dragdiags} describes the scattering of an impurity with a particle, with momentum transfer $\q$, and the bottom diagram the scattering with a hole, with momentum transfer $-\q$. Therefore, the net momentum transfer and drag vanish in the particle-hole symmetric case \cite{kamenev1995coulomb, PhysRevB.76.081401,RevModPhys.88.025003}, as will be seen explictly below. 
The remaining evaluation of the diagrams is straightforward (see App.\ \ref{appdragsec}). We obtain 
\begin{widetext}
\begin{align}
\label{maineqdragcont}
&\sigma_{\downarrow\uparrow} = -2 g^2 n_\downarrow \int \frac{d\k}{(2\pi)^2} \frac{d\q}{(2\pi)^2} \  \text{Im} \left\{ J^y_{\uparrow,12}(\k) W^{22}(\k - \q, \q) W^{21}(\k,-\q)\right\}  \frac{q_x}{M} \frac{1}{\left(\eone(\k) - \etwo(\k)\right)^2}  \left(d(\k,\q) + c(\k,\q) \right) , \\
\label{cddef}
& d(\k,\q) = \frac{2\epsilon_{\uparrow,1}(\k) - \epsilon_{\uparrow,2}(\k) - \epsilon_{\uparrow,2}(\k - \q) - \epsilon_\downarrow(\q)}{\left(\eone(\k)- \etwo(\k-\q)-\epsilon_\downarrow(\q)\right)^3} \ , \quad 
c(\k,\q) = \frac{2 \etwo(\k) - \eone(\k) - \eone(\k-\q) + \epsilon_\downarrow(\q)}{(\eone(\k-\q) - \etwo(\k) - \epsilon_\downarrow(\q))^3} \ . 
\end{align}
\end{widetext}
Here, $c,d$ represent the contributions of the ``direct'' (top in Fig.\ \ref{dragdiags}) and ``crossed'' (bottom) diagrams. When flipping $d_2 \rightarrow - d_2$, we have $\epsilon_1 \rightarrow - \epsilon_2$ and vice versa, thus $\sigmad$ is antisymmetric in $d_2$. In particular, it vanishes in the particle-hole symmetric case, $d_2 = 0$. Numerical evaluation of Eq.\ \eqref{maineqdragcont} as function of $d_2$ is shown in Fig.\ \ref{contplotnums}(a). Let us point out that the complete cancellation of $\sigma_{\downarrow\uparrow}$ at $d_2 = 0$ only occurs to second order, $\mathcal{O}(g^2)$, and is not expected in higher order, as can be shown explicitly for the Haldane model (see below). 

In Fig.\ \ref{contplotnums}(b), $\sigmad$ is depicted as function of $m$ for non-zero $d_2$, tuning the majority system from the trivial phase with $\mathcal{C} = 0$ to a non-trivial one, $\mathcal{C} = 1$. While $\sigmad$ exhibits a clear jump when the majority particles undergo a topological phase transition (see next section), it is neither constant in the non-trivial phase, nor does it vanish in the trivial phase: For the majority particles, time-reversal symmetry is broken everywhere in the phase diagram, but for $\mathcal{C} = 0$ the transconductivity contributions of the ``Dirac'' and ``spectator'' fermions cancel exactly, as long as the chemical potential is in the gap and the lower majority band is completely filled. In the case of the gapless impurity band, such cancellation is not guaranteed, and the  impurity Hall drag therefore does not vanish in the non-trivial phase. 

\section{The jump across the phase transition for the continuum model}
\label{jumpsec}

Another salient feature of Fig.\ \ref{contplotnums}(b) is the discontinuous change of the drag transconductivity which occurs upon crossing the topological phase boundary $m=0$. This jump can be understood as arising from a singular contribution of Dirac fermions: When the gap closes, the Dirac part of the majority Berry curvature ($\propto m$ in Eq.\ \eqref{Cexp}) evolves into a delta-function, $\text{sign}(m) \delta^{(2)}(\k)$ -- compare also Fig.\ \ref{contbands}. In contrast, the part corresponding to the spectator fermions ($\propto d_1$ in Eq.\ \eqref{Cexp}) is smooth across the transition. In the expression for the impurity drag \eqref{maineqdragcont}, a singular Dirac contribution $\propto \text{sign}(m) \delta^{(2)}(\k)$ arises as well. This singular contribution changes sign across the transition, and so induces the jump $\Delta \sigmad$  in the Hall drag, with a sign determined by the change in Chern number $\Delta\mathcal{C}$. To extract $\Delta \sigmad$ we can set $\k = 0$ in all parts of Eq.\ \eqref{maineqdragcont} which are non-singular as $\k \rightarrow 0$. As detailed in App.\ \ref{contjumpapp}, in this way we obtain

\begin{widetext}
\begin{align}
\label{sigmaDiracfirst} &\sigma_{\downarrow\uparrow,\text{Dirac}} =  \\&\notag  g^2 n_\downarrow  \int \frac{d\k}{(2\pi)^2} \frac{d\q}{(2\pi)^2} \  \frac{\text{Im} \left\{ J^y_{\uparrow,\text{Dirac},12}(\k)  J^x_{\uparrow, \text{Dirac},21}(\k)\right\}}{\left(\eone(\k) - \etwo(\k)\right)^2}  \frac{q_x^2}{M q\sqrt{1+(d_1 q)^2}}\left( \frac{1}{(\epsilon_\downarrow(\q) + \etwo(\q))^2} -  \frac{1}{(\epsilon_\downarrow(\q) - \eone(\q))^2} \right) \ , 
\end{align}
\end{widetext}
where $J^{x/y}_{\uparrow,\text{Dirac}}(\k)$ represents the majority current carried by the Dirac (i.e., not the spectator) fermions. Compared to Eq.\ \eqref{maineqdragcont} the $\k$ and $\q$ integrals in Eq.\ \eqref{sigmaDiracfirst} have factorized. The $\k$ integral, which simplifies to an integral over a delta function as $m \rightarrow 0$, is nothing but the Chern number contribution of the Dirac fermions, cf.\ Eq.\ \eqref{sigmaCrel}. It evaluates to $(1/8\pi ) \text{sgn}(m)$. Performing the remaining $\q$ integral, one finds 
\begin{align} 
\label{diracfinal}
\sigma_{\downarrow\uparrow,\text{Dirac}} = -  \frac{g^2 n_\downarrow}{(2\pi)^4} \cdot \frac{4\pi^2 d_2 M\cdot\text{sign}(m)}{1+ 4M (|d_1| + (d_1^2 - d_2^2) M)} \   .
\end{align}

\begin{figure}
\centering
\includegraphics[width=.95\columnwidth]{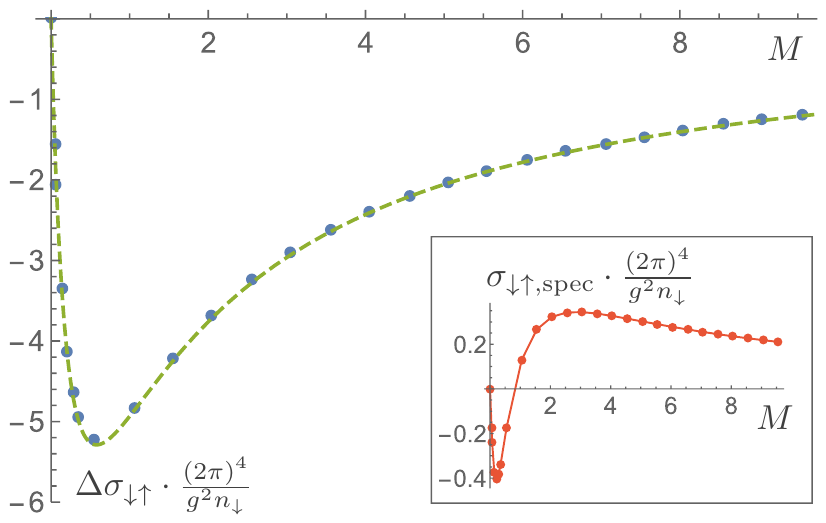}
\caption{Jump of the Hall drag $\Delta\sigmad$ in the continuum model as function of $M$, with $d_1 = -1, d_2 = 0.5$. The dashed line corresponds to Eq.\ \eqref{deltasigmaC}, points are computed numerically by evaluating Eq.\ \eqref{maineqdragcont} at two points $m = \pm 0.001$ close to the phase boundary. Numerical errors are of the order of the points size. \textit{Inset}: The smooth contribution of the spectator fermions, obtained numerically from Eq.\  \eqref{maineqdragcont}.}
\label{contjump}
\end{figure}

Defining $\Delta\sigmad$ as the jump of Hall drag when going from the trivial to the topological phase, with change in Chern number $\Delta \mathcal{C}$, Eq.\ \eqref{diracfinal} leads to the final result: 
\begin{align}
\label{deltasigmaC}
 \Delta\sigmad = \Delta \mathcal{C}\cdot \frac{g^2 n_\downarrow}{(2\pi)^4}\left(- \frac{8\pi^2 d_2 M}{1+ 4M (|d_1| + (d_1^2 - d_2^2) M)}\right) \   .
\end{align}
As a check, in Fig.\ \ref{contjump} this formula is compared with a numerical evaluation of the jump from Eq.\ \eqref{maineqdragcont} as function of the impurity mass $M$, yielding excellent agreement. Note that both Hall drag and jump will vanish in the limits $M \to 0 \ \text{or} \ M \to \infty$: In the former limit, the impurity cannot interact efficiently with the majority particles due to the large kinetic energy cost, while in the latter the impurity is immobile and cannot be dragged. 

While the Dirac part of the Hall drag, $\sigma_{\downarrow\uparrow,\text{Dirac}}$, changes sign at the transition, there is also a small smooth background contribution from the spectator fermions, to be denoted $\sigma_{\downarrow\uparrow,\text{spec}}$. This contribution can be extracted numerically from Eq.\ \eqref{maineqdragcont} as  
\begin{align} 
\sigma_{\downarrow\uparrow,\text{spec}}= \frac{1}{2}\left[\sigmad(m = 0^+) + \sigmad(m = 0^-) \right]\ , 
\end{align}  see the inset to Fig.\ \ref{contjump}. 

We note in passing  that the jump of $\sigmad$ is reminiscient of the recently shown \cite{PhysRevLett.126.076603} change of sign in the Hall \textit{coefficient} for a single-particle gapless Dirac cone upon variation of the particle density.

\section{Drag and jump in the Haldane lattice model}
\label{Haldanemodel}

\begin{figure*}
\centering
\includegraphics[width=\textwidth]{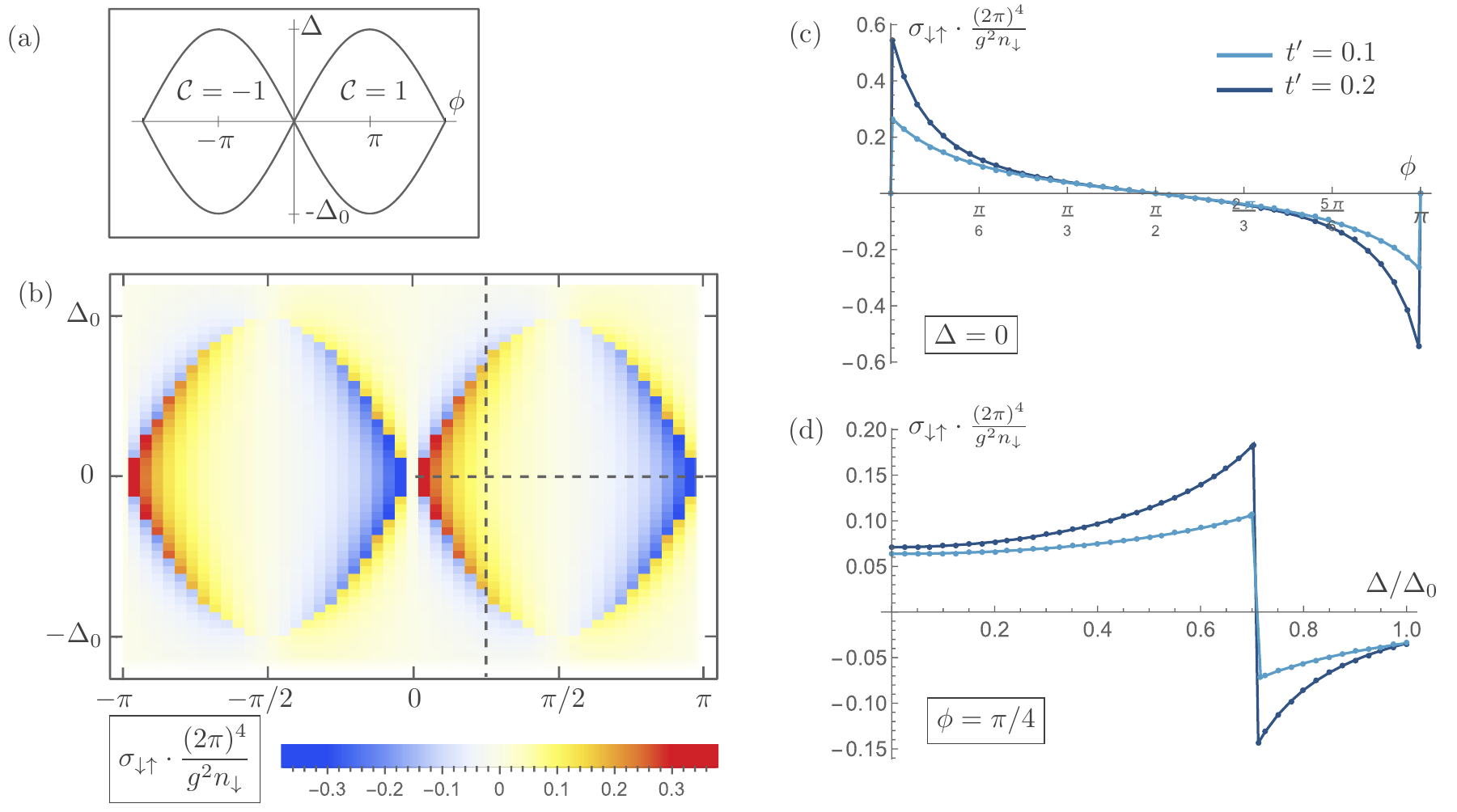}
\caption{Impurity Hall drag $\sigmad$ in the Haldane model. (a) Majority phase diagram. $\Delta_0 = 6\sqrt{3}t^\prime$ is the value of $\Delta$ where the phase transition occurs for $\phi = \pi/2$. (b) $\sigmad$ from numerical evaluation of Eq.\ \eqref{haldanedragformula} for $t^\prime = 0.2$.  Cuts through the phase diagram along the dashed lines are shown in the next panels. (c) $\sigmad$ as function of $\phi$ for $\Delta = 0$ and two values of $t^\prime$. (d) $\sigmad$ as function of $\Delta$ for $\phi = \pi/4$ and same two values of $t^\prime$. The abscissa is rescaled by $\Delta_0(t^\prime)$. }
\label{haldane_allfig}
\end{figure*}

The general behaviour of $\sigma_{\downarrow\uparrow}$ to leading order, $\mathcal{O}(g^2)$, is not limited to the continuum model \eqref{contham}, but will hold in other Chern insulators as well. As another example, we consider a situation \cite{PhysRevB.99.081105, *PhysRevB.102.119903} where the majority particles are described by the  Haldane model on the honeycomb lattice  \cite{PhysRevLett.61.2015}, with Hamiltonian 
\begin{align}
\label{Haldaneham}
H_\uparrow(\k)  &= \sum_{i = 0}^3 \psi_\uparrow^\dagger(\k) \left(h_i(\k)  \sigma_i \right) \psi_\uparrow(\k)\ ,\\ \notag   \psi_\uparrow(\k) &= \left(c_{\uparrow,A}(\k), c_{\uparrow,B} (\k)\right)^T, \ \quad \k_i = \k \cdot \boldsymbol{u}_i\ , \\ \notag 
 h_0(\k) &= -2t^\prime \cos(\phi) \left[ \cos(\k_1 - \k_2) + \cos(\k_1) + \cos(\k_2) \right] \ ,  \\ \notag h_1(\k) &= - \left[1 + \cos(\k_1) + \cos(\k_2)\right]\ , \\  \notag  h_2(\k) &= -\left[(\sin(\k_1) + \sin(\k_2)\right], \\ \notag h_3(\k) &= 
 \\& \notag \hspace{-2em} \Delta/2  + 2t^\prime \sin(\phi) \left[ \sin(\k_1 - \k_2) + \sin(\k_2) - \sin(\k_1)\right] \ ,  \end{align} 
 where $ \ \uone = (3/2, \sqrt{3}/2)^T, \  \utwo= (3/2, - \sqrt{3}/2)^T$, and the lattice constant and nearest neighbour hopping amplitude are set to 1. The reciprocal lattice vectors are given by $\boldsymbol{b}_1 = (2\pi/3,2\pi/\sqrt{3})^T, \ \boldsymbol{b}_2 = (-2\pi/3,2\pi/\sqrt{3})^T$. The model is parametrized by the next-nearest-neighbour hopping $t^\prime$, the angle $\phi$ quantifying the time-reversal symmetry breaking, and the sublattice potential offset $\Delta$. 
For given values of $t^\prime, \phi, \Delta$, the majority chemical potential is implicitly placed in the gap (its precise value is irrelevant). The well-known topological phase diagram of the Haldane model is shown in Fig.\ \ref{haldane_allfig}(a). 
 
The impurity particles are governed by the tight-binding model for graphene (i.e., $t^\prime = \Delta = 0$), with the chemical potential at the bottom of the lower band \cite{PhysRevB.99.081105} {by setting $h_0(\k) = 3$}. The impurity-majority interaction, Eq.\ \eqref{Hint}, is straigthforwardly modified to account for the impurity multi-band structure.

The Hall drag $\sigmad$ can then be derived in analogy to the continuum model, see App.~\ref{Haldaneapp} for details; the only minor change is the appearance of diagonalizing unitary matrices $U_\downarrow(\q)$ for the impurity. Numerical evaluation of $\sigmad$ is presented in Figs.\ \ref{haldane_allfig}(b)--(d). Now, the particle-hole symmetric case where $\epsilon_1 = - \epsilon_2$ corresponds to $\phi = \pm  \pi/2$, and $\sigmad$ vanishes accordingly \cite{PhysRevB.102.119903}. Furthermore, one can easily demonstrate  the symmetry $\sigmad(\phi) = - \sigmad(\pi - \phi)$, see App.\ \ref{Haldaneapp} below Eq.\  \eqref{haldanedragformula}. This symmetry is readily seen in Fig.\ \ref{haldane_allfig}(c), which shows a cut through the phase diagrams for fixed $\Delta = 0$. Combined with the symmetry $\sigmad(\phi) = - \sigmad(-\phi)$ inherited from the Haldane model, this gives the Hall drag a periodicity
\begin{align}
 \sigmad(\phi) = \sigmad(\phi + \pi)\ , 
\end{align} 
apparent in Fig.\ \ref{haldane_allfig}(b). This remarkable manifestation of particle-hole antisymmetry is in stark contrast to the pure majority case, where the Chern number only has the trivial periodicity $\mathcal{C}(\phi) = \mathcal{C}(\phi + 2\pi)$, see Fig.\  \ref{haldane_allfig}(a).

At the special particle-hole symmetric parameter points, $\phi = \pm \pi/2, \Delta =0$, one can also get insight into the behavior of $\sigmad$ to higher order in $g$ (see App.\ \ref{antapp}):  employing a particle-hole transformation which also exchanges band indices of the majority particles, it can be shown that at these points the Hall drag is antisymmetric in $g$ to all orders. So while there is no $\mathcal{O}(g)$ contribution, and the leading order, $\mathcal{O}(g^2)$, must vanish, at order $\mathcal{O}(g^3)$ the Hall drag will be nonzero.  \\

\begin{figure}
\centering
\includegraphics[width=.95\columnwidth]{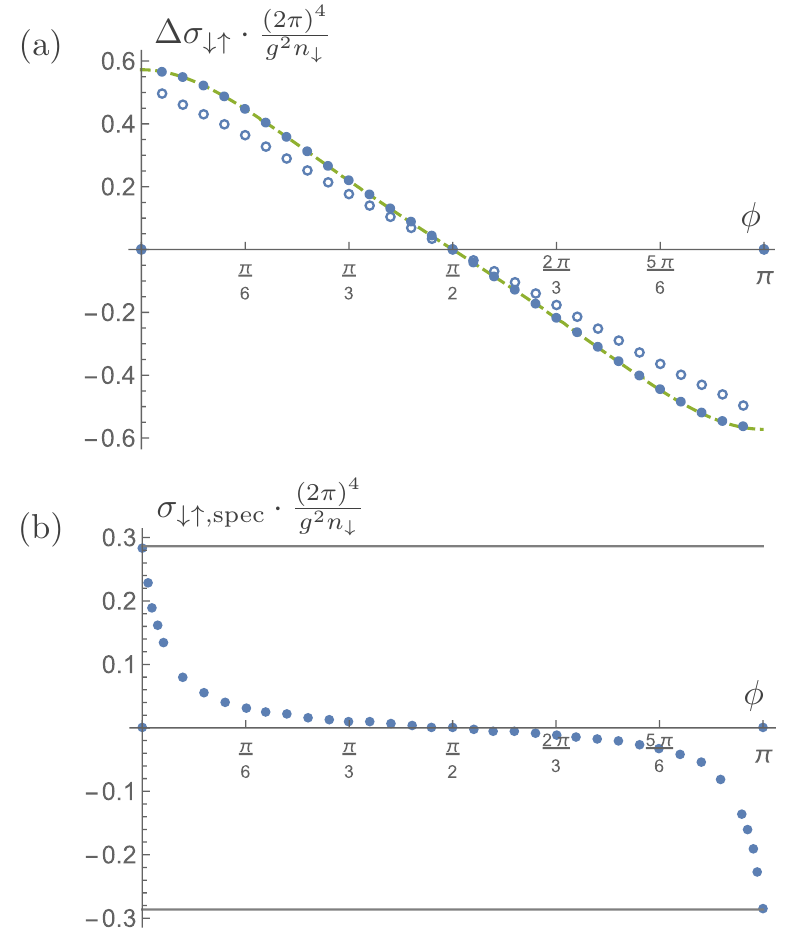}
\caption{(a) Jump of the Hall drag $\Delta\sigmad$ in the Haldane model as function of $\phi$, with $t^\prime = 0.2$ and $\Delta = \Delta_c$ tuned to the transition line.  The dashed line corresponds to formula \eqref{haldanejumpmain}, points are computed numerically by evaluating Eq.\ \eqref{haldanedragformula} at two points close to the phase boundary, with $\Delta = \Delta_c \pm 0.001$ (filled circles). For comparison, a numerical evaluation with $\Delta = \Delta_c \pm 0.1$ is also shown (empty circles), which yields qualitative agreement only. (b) Smooth contribution from spectator fermions, obtained numerically from \eqref{haldanedragformula}. Horizontal lines correspond to $\Delta\sigmad(\phi = 0^+, \pi^-)/2.$}
\label{Haldanejump}
\end{figure}

In the numerics, the jump of $\sigmad$ across the topological phase transition is again  prominent, and clearly delineates the topological phases of the parent Haldane model. Its origin is analogous to the continuum model -- it comes from a sign-changing contribution of Dirac fermions, which becomes singular upon gap closing. The only significant difference is that there are now two Dirac cones in the problem, but except at the special points $\phi = 0,\pi$, the gap closes at only one of them. In the language employed for the continuum model, states near the Dirac cone with open gap count as spectator fermions. A detailed analysis of the jump leads to (see App.\ \ref{Haldaneapp})
\begin{align} 
\label{haldanejumpmain}
\Delta\sigmad =  \Delta \mathcal{C} \cdot  \frac{g^2 n_\downarrow}{(2\pi)^4} \cdot  f(t^\prime, \phi) , 
\end{align}
where $f(t^\prime, \phi)$ is a numerical function defined in Eq.\ \eqref{sigmadirachald}. It involves the remaining $\q$ integral, which is difficult to evaluate analytically in the lattice case.  In Fig.\ \ref{Haldanejump}(a), $\Delta\sigmad$ is depicted as a function of $\phi$. It is maximal as $\phi \rightarrow 0^+, \pi^-$, where the particle-hole asymmetry of the dispersion (away from the Dirac points) is largest. Again, the jump occurs on top of a smooth background contribution from the spectator fermions, presented in Fig.\ \ref{Haldanejump}(b). It too is maximal as $\phi \rightarrow 0^+, \pi^-$, approaching $1/2\Delta\sigmad$: close to these angles, the spectator contribution is almost fully determined by the second Dirac cone, which has a very small gap. Accordingly, the values of the sign-changing drag contribution, $\sigma_{\downarrow\uparrow, \text{Dirac}}$, and the almost Dirac-like background contribution are the same.

\section{Measurement of the Hall drag} 
\label{dichroismsec}

We now discuss how to detect $\sigmad$ experimentally. In a solid state system, the total transversal conductivity $\sigma_{xy, \text{tot}}$ is an easily accessible quantity, typically obtained from a resistivity measurement. Since the majority particles form a Chern insulator, their contribution to $\sigma_{xy, \text{tot}}$ is quantized, and the Hall drag contribution $\sigmad$ can in principle be read off by subtracting this quantized value from $\sigma_{xy, \text{tot}}$. In practice, however, it may be necessary to use the specific parameter dependence of $\sigmad$ to separate it from $\sigma_{xy, \text{tot}}$. $\sigmad$ can for example be obtained as the contribution to $\sigma_{xy, \text{tot}}$ proportional to the impurity density $n_\downarrow$, or by subtracting measurements of  $\sigma_{xy, \text{tot}}$ at two particle-hole inverted points of the phase diagram. 

Chern insulators have also been successfully realized in ultracold gas systems. Here, an established technique for measuring topological quantum numbers \cite{Aidelsburger2015,PhysRevA.102.063316} is the in-situ observation of the center of mass displacement of the atomic cloud upon the action of an external force. In the present polaron context, this measurement would have to be performed in a state-dependent manner to extract the Hall drag. In addition, one could conduct either a state-dependent time-of-flight measurement \cite{PhysRevLett.109.095302, Wu83}, or Raman spectroscopy (as recently implemented for polarons \cite{PhysRevX.10.041019}), to infer the in-trap momentum distribution of the impurity, in view of evaluating the current response of the impurity to an applied force.

%

All these transport experiments would extract the Hall drag from the linear current response to an external, linearly polarized electric field, which is the standard point of view. However, recent theoretical works have shown \cite{tran2017probing, PhysRevA.97.061602, PhysRevLett.122.166801,PhysRevResearch.2.033385,PhysRevB.103.035114} that topological invariants can also be obtained from a measurement of excitation rates to second order in the amplitudes of circularly polarized fields, which was verified in the experiment of Ref.\ \cite{asteria2019measuring}. For the Hall drag $\sigmad$, a relation to an impurity excitation rate can be established as well, as we now show. Measuring such excitation rates may be a simpler route to detect $\sigmad$ experimentally, in both ultracold gas and solid state systems.

To set the stage, we first rephrase the results of Ref.\ \cite{tran2017probing} for the majority sector (non-interacting Chern insulator). The particles are coupled to external left or right circular polarized electrical fields: 
\begin{align}
\textbf{E}_{\pm}(t) = 2E \left(\cos(\omega t),\ \pm \sin(\omega t)\right)^T \ , 
\end{align}
with $\omega$ a fixed drive frequency. 
In the temporal gauge, the time-dependent light-matter Hamiltonian reads
\begin{align}
\label{Hpmup}
&H_{\uparrow,\pm}(t) = \frac{2 E}{\omega} \left( \hat{J}^x_\uparrow  \sin(\omega t) \mp \hat{J}^y_{\uparrow} \cos(\omega t) \right) \ . 
\end{align}
When this perturbation is switched on, particles are excited from the lower to the upper band. One can define the associated depletion rates of initially occupied states with momentum $\k$, $\Gamma_{\uparrow,\pm}(\k,\omega)$, which depend on the polarization of the driving field (``circular dichroism''). In Ref.\ 
\cite{tran2017probing}, these rates are obtained from Fermi's Golden Rule. Let $\Delta\Gamma_\uparrow(\omega)$ be the difference in total depletion rates for a fixed frequency $\omega$, $\Delta \Gamma_\uparrow(\omega) \equiv 1/2 \sum_\k (\Gamma_{\uparrow,+}(\k,\omega) - \Gamma_{\uparrow,-}(\k,\omega))$. Then the Chern number $\mathcal{C}$ follows the simple relation \footnote{Note that we use a different sign convention for $\mathcal{C}$ than Ref.\  \cite{tran2017probing}}: 
\begin{align} 
\label{dichrobasic}
A_0 E^2 \mathcal{C} = - \int_{0}^\infty \ d\omega \Delta \Gamma_\uparrow (\omega) \ . 
\end{align}
This integration has to be understood as an average of $\Delta\Gamma_\uparrow(\omega)$ over different drive frequencies, obtained by repeating the experiment many times \cite{asteria2019measuring}.

\begin{figure}[b]
\centering
\includegraphics[width=.75\columnwidth]{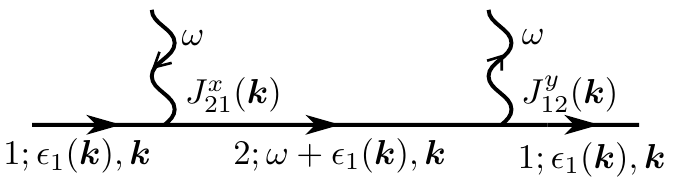}
\caption{On-shell self-energy diagram. Incoming and outgoing fermion lines represent particles from the lower band, the intermediate line a particle from the upper band, and the wiggly lines the circularly polarized electrical fields. The Feynman rules are explained in App.\ \ref{dichroapp}.}
\label{rate_nonint}
\end{figure}

For our purposes here, it is useful to rederive the result \eqref{dichrobasic} from diagrammatic perturbation theory. This is achieved by relating the depletion rate to the on-shell retarded self-energy as 
\begin{align}
\Gamma_{\pm, \uparrow}(\k,\omega) = - 2 \text{Im}\left[\Sigma_\pm(\epsilon_{\uparrow,1}(\k), \k; \omega) \right] \ . 
\end{align}
In turn, the self-energy to second order in $H_{\uparrow,\pm}$ can be represented by the Feynman diagram of Fig.\ \ref{rate_nonint}, plus the diagram with the $\hat{J}^x_\uparrow, \hat{J}_\uparrow^y$ vertices interchanged. The necessary Feynman rules in energy-momentum space are easily derived from $H_{\uparrow, \pm}$, and are detailed in App.\ \ref{dichroapp}. There are also processes involving $(\hat{J}^x_\uparrow)^2, (\hat{J}^y_\uparrow)^2$, but they cancel in $\Delta \Gamma_\uparrow(\omega)$. Working directly in the real frequency space for convenience, $\Delta \Gamma_\uparrow(\omega)$ can then be directly written down as 
\begin{align}
&\Delta\Gamma_\uparrow(\omega) \notag =  - \sum_\k \text{Im} \left[ \Sigma_+(\epsilon_1(\k), \k; \omega)- \Sigma_{-}(\epsilon_1(\k), \k; \omega)\right]= \\ \notag & -\sum_\k \frac{E^2}{\omega^2}\text{Im}\bigg[
\left(2iJ_{\uparrow,21}^x(\k)J^y_{\uparrow,12}(\k) - 2iJ_{\uparrow,21}^y(\k) J_{\uparrow,12}^x(\k) \right)   \\ &  \qquad \qquad \qquad \times \frac{1}{\omega + \epsilon_{\uparrow,1}(\k) -\epsilon_{\uparrow,2}(\k) + i0^+} \bigg] \ . 
\label{DeltaGammasome}
\end{align} 
Integrating over $\omega$, we find: 
\begin{align}
\label{majratemycomp}
&\int_0^\infty d\omega \Delta\Gamma_\uparrow(\omega)  =  \\ \notag &\frac{4\pi E^2 A_0}{(2\pi)^2} \int d\k \int_0^\infty d\omega \frac{\delta(\omega - (\epsilon_{\uparrow,2}(\k) -\epsilon_{\uparrow,1}(\k)))}{\omega^2} \\  \notag &   \times  \text{Im} \left[ J_{\uparrow,12}^x(\k) J_{\uparrow,21}^y(\k) \right] \overset{\eqref{sigmaCrel}}{=} - A_0 E^2 \mathcal{C}\  , 
\end{align}
in agreement with Eq.\ \eqref{dichrobasic}. 

To summarize, we have related the majority Chern number to the differential depletion rate of filled states from the lower band when the system is subjected to a circular perturbation. We can now extend this idea to the impurity case. We consider our previous interacting majority-impurity setup, with a small number of impurities prepared in the lower band, 
and couple both majority and impurity particles to the circular fields. On their own, the impurities would not experience a differential depletion because of the time reversal invariance of the impurity Hamiltonian. Only due the interaction with the majority particles such differential depletion will set in, corresponding to occupation of higher momentum states. Note that, for strong impurity-majority interactions, it will rather be polaronic (dressed impurity) states which are depleted. For weak coupling, however, such band-dressing effects can be neglected (to order $\mathcal{O}(g^2)$), and we can think in terms of bare impurities in lieu of polarons. In technical terms, our Feynman diagrams will not contain any impurity self-energy insertions.

\begin{figure*}
\centering
\includegraphics[width=.95\textwidth]{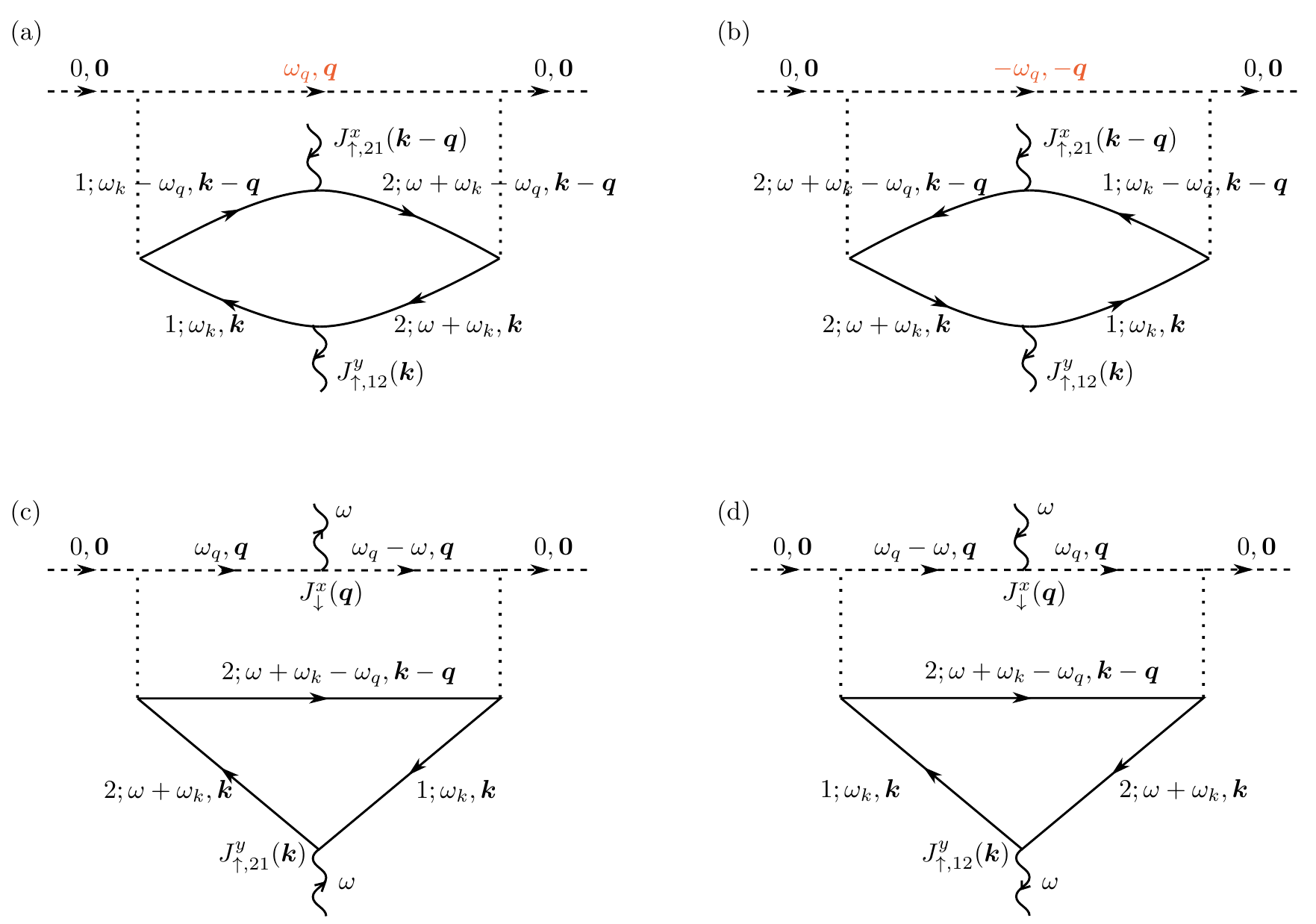}\caption{Non-vanishing contributions to the impurity depletion rate $\Gamma_{\downarrow,\pm}(\0,\omega)$. Panels (a), (b): Diagrams  not related to the drag, which are particle-hole symmetric. Panels (c), (d): Diagrams related to the drag. These  two diagrams differ in the orientation of the field lines and the band index structure of the majority particles.  }
\label{imp_rate}
\end{figure*}

Let us couple the impurities to the circular fields in the same way as the majority particles, Eq.\ \eqref{Hpmup}. We consider the depletion rate of the filled impurity state with vanishing momentum $\Gamma_{\downarrow,\pm}(\0,\omega) \equiv \Gamma_{\downarrow,\pm}(\omega) $, which is of most interest when the impurity density is small. Since non-vanishing contributions to $\Delta \Gamma_\downarrow(\omega)$ must involve majority scattering, to order $\mathcal{O}(g^2)$ there are two classes of relevant diagrams; representative diagrams are shown in Fig. \ref{imp_rate}. 

Consider first the two diagrams  \ref{imp_rate}(a), \ref{imp_rate}(b) in the top row of the Figure. These diagrams describe processes where only the majority particles are excited by the external fields. Since they do not involve an impurity current, they are not related to the drag. Two additional  diagrams where the direction of the external field lines is inverted can be drawn as well. 

The structural difference between Fig.\ \ref{imp_rate}(a) and \ref{imp_rate}(b) is the orientation of the majority lines, which maps to an inverted energy-momentum transfer on the impurity (marked red). Thus, similar to the drag diagrams of Fig.\ \ref{dragdiags}, the diagrams  are related by particle-hole symmetry. However, the contributions of these diagrams add up rather than cancel, since they do not contain an impurity current operator, $J_\downarrow(\q)$, which is odd in $\q$. Therefore, as can be verified by a straightforward evaluation (cf.\ App.\ \ref{dichroapp}, Eq.\ \eqref{extraEq}), the total contribution $\Delta\Gamma_{\downarrow,\text{ph}}$ of these diagrams obeys $\Delta \Gamma_{\downarrow,\text{ph}}(\phi) = \Delta \Gamma_{\downarrow,\text{ph}}(\pi - \phi)$ for the Haldane and $\Delta \Gamma_{\downarrow,\text{ph}}(d_2) = \Delta \Gamma_{\downarrow,\text{ph}}(-d_2)$ for the continuum model. As a result, in an experiment these processes can be projected out by subtracting $\Delta \Gamma_{\downarrow,\text{ph}}(\phi) - \Delta \Gamma_{\downarrow,\text{ph}}(\pi - \phi)$, which leaves out only the antisymmetric drag contribution. Another way to separate $\Delta\Gamma_{\downarrow,\text{ph}}$  from the drag is to have a different coupling constant between external field and impurities, which is feasible in the ultracold gas setup where the circular perturbation can for example be implemented by lattice shaking \cite{PhysRevLett.115.073002, asteria2019measuring}. Since $\Delta\Gamma_{\downarrow,\text{ph}}$ is independent of the coupling to the impurities, it can again be eliminated by subtracting measurements obtained for two different impurity couplings.

Let us assume either such elimination implicitly, and move on to the two diagrams of  Fig.\ \ref{imp_rate}(c), \ref{imp_rate}(d).  In essence, they correspond to the drag transconductivity diagram of Fig.\ \ref{dragdiags} (top), with the central (red) impurity line cut. The two other diagrams in this class have  crossed interaction lines, akin to the  ``crossed'' diagrams of Fig.\ \ref{dragdiags} (bottom). Evaluation of these four diagrams is straightforward, see App.\ \ref{dichroapp}. Summation over the filled impurity states simply yields:
\begin{align}
\sum_{\p, \text{filled}} \Gamma_{\downarrow,\pm}(\p,\omega) \simeq \sum_{\p, \text{filled}} \Gamma_{\downarrow,\pm}(\omega) = A_0 n_\downarrow \Gamma_{\downarrow,\pm}(\omega)\ . 
\end{align}
For the integrated differential depletion rate, one then finds 
\begin{align} 
\int_0^\infty d\omega \Delta \Gamma_{\downarrow, xy}(\omega) = 2\pi A_0 E^2 \sigmad \ , 
\end{align}
as naively expected from Eq.\ \eqref{majratemycomp}. However, the impurity depletion rate also receives contribution from processes involving the currents $\hat{J}_\downarrow^y$, $\hat{J}_\uparrow^x$. Per the Feynman rules (cf.\ App.\ \ref{dichroapp}), these diagrams come with a relative minus sign, and then yield a factor of two for the total differential rate, since $\sigma_{xy,\downarrow\uparrow} = - \sigma_{yx, \downarrow\uparrow}$ for both the continuum and the Haldane model, as one can check easily. Modulo the antisymmetrization discussed above, we therefore have
\begin{align}
\label{dichrofinal}
\sigmad= \frac{1}{4\pi A_0 E^2} \int_0^\infty d\omega \Delta \Gamma_{\downarrow}(\omega) \  . 
\end{align}

This result can also be rephrased in terms of excitation instead of depletion rates. Since the impurities are initially prepared at the bottom of the band, one can write 
\begin{align}
\label{absorption rates}
\int_0^\infty d\omega \Delta \Gamma_\downarrow(\omega) = \sum_{\q>0} \  \int_0^\infty d\omega  \Delta\Gamma_{\downarrow, \text{exc}}(\q, \omega) \ , 
\end{align}
meaning that the impurities are excited \textit{into} states with higher momentum which are initially empty. These $\q$-states correspond to the intermediate impurity lines in Fig.\ \ref{imp_rate}. Via Eq.\ \eqref{dichrofinal} we can then define a $\q$-resolved impurity drag as 
\begin{align}
\sigmad \equiv \sum_{\q>0} \sigmad(\q) \ . 
\end{align}

 This provides an alternative view on, say, the topological jump $\Delta\sigmad$. For the Haldane model, it can be phrased as $\Delta\sigmad = \Delta \mathcal{C} \int d\q f_\text{jump}(\q)$, where $f_{\text{jump}}(\q)$ is a known function, see Eqs.\ \eqref{haldanejumpmain}, \eqref{sigmadirachald}. If the excitation rates defined in Eq.\ \eqref{absorption rates} can be experimentally detected in $\q$-resolved fashion (for instance with band mapping techniques \cite{PhysRevLett.94.080403, Tarruell2012, Jotzu2014}), so can the $\q$-resolved impurity drag $\sigmad(\q)$. Measuring $\sigmad(\q)$ at two points in the phase diagram close to the topological boundary then gives direct access to $f_\text{jump}(\q)$. Taken the other way around, supposing that $f_{\text{jump}}(\q)$ is known for the model realized in the experiment, at each $\q$-point an independent estimate of the change in Chern number across the phase transition $\Delta \mathcal{C}$ is possible. 

\section{Conclusions}
\label{conclusionsec}

In this work we have studied to which extent a topologically trivial impurity can be Hall-dragged by majority excitations in a Chern insulator, looking at two different models in a controlled perturbative setting. Since the impurity Hall drag is sensitive to the dispersion of the majority particles and holes, there is no one-to-one correspondence to the Chern number; nevertheless, the change in Chern number across a topological transition is clearly reflected by a discontinuous jump in the drag transconductivity $\sigmad$. This jump arises from the integrated singular Berry curvature of the majority fermions.The transconductivity can be extracted either from transport experiments, or from a measurement of impurity excitation rates upon driving the system by a circularly polarized field. 

A worthwhile goal for future study is the extension to the strong-coupling limit, in particular the analysis of impurity-majority bound state formation. These bound states may have rather rich physics: They could inherit the topological characteristics of the majority particles \cite{Grusdt2016, PhysRevB.100.075126}, have opposite chirality as found for Haldane model in the two-body limit \cite{PhysRevA.97.013637}, or even be topological when the single-particle state are trivial \cite{PhysRevA.101.023620, Olekhno2020,PhysRevResearch.2.013348}.

\section*{Acknowledgments}

We thank A.\ Kamenev for helpful discussions. D.P.\ acknowledges funding by the Deutsche Forschungsgemeinschaft (DFG, German Research Foundation) under Germany's Excellence Strategy -- EXC-2111 -- 390814868, and is particularly grateful to the Max-Planck-Institute for the Physics of Complex Systems Dresden (MPIPKS) for hospitality during the intermediate stage of this project. N.G.\ has been supported by the FRS-FNRS (Belgium) and the ERC Starting Grant TopoCold. P.M.\ has been supported by the Spanish MINECO (FIS2017-84114-C2-1- P), and EU FEDER Quantumcat. G.M.B.\ acknowledges support from the Independent Research Fund Denmark-Natural Sciences via Grant No.\ DFF-8021-00233B, and US Army CCDC Atlantic Basic and Applied Research via grant W911NF-19-1-0403. M.G.\ has been supported by the Israel Science Foundation (Grant No.\ 227/15) and the US-Israel Binational Science Foundation (Grant No.\ 2016224).
\\


\label{conclusionsec}

\appendix 


			
\section{Basis rotation}
\label{BasisrotSec}

To evaluate the conductivities, it is convenient to work in the diagonal band basis, introducing a diagonalizing unitary matrix $U_\uparrow(\k)$
\begin{align}
\label{rotdef}
&U_\uparrow^\dagger(\k) H_\uparrow(\k) U_\uparrow(\k) = \text{diag}(\epsilon_{\uparrow,1}(\k), \epsilon_{\uparrow,2}(\k))  \\
\label{Uall}
&U_\uparrow(\k) = \begin{pmatrix} U_{\uparrow,A1}(\k) & U_{\uparrow,A2}(\k) \\  U_{\uparrow,B1}(\k) & U_{\uparrow,B2}(\k)
\end{pmatrix},  \\ & \notag  
U_{\uparrow,A1}(\k) = \frac{h_3(\k) - h(\k)}{\sqrt{2h(\k) (h(\k) - h_3(\k))}} \\ \notag & U_{\uparrow,A2}(\k) = \frac{h_3(\k) + h(\k)}{\sqrt{2h(\k) (h(\k) + h_3(\k))}}
  \\ \notag & U_{\uparrow,B1}(\k) = \frac{h_1(\k) + ih_2(\k)}{\sqrt{2h(\k) (h(\k) - h_3(\k))}}, \\ \notag & U_{\uparrow,B2}(\k) = \frac{h_1(\k) + ih_2(\k)}{\sqrt{2h(\k) (h(\k) + h_3(\k))}} \ , 
\end{align}
where $A,B$ refer to the sublattice-  and $1,2$ to the diagonal band basis. The same expressions apply for the Haldane model as well. 

In the band basis, the second-quantized current operator is given by
\begin{align}
\hat{J}^{x/y}_\uparrow &= \sum_\k c^\dagger_{\uparrow,\alpha}(\k) J^{x/y}_{\uparrow,\alpha\beta}(\k) c_{\uparrow,\beta}(\k), 
\end{align}
with matrix elements
\begin{align}
J_\uparrow^{x}(\k) &= U_\uparrow^\dagger(\k) J_\uparrow^{x,0}(\k) U_\uparrow(\k) \\ 
J_\uparrow^{x,0}(\k) &= \frac{\partial H_\uparrow(\k)}{\partial k_x} =  \sigma_x + 2k_x (d_1 \sigma_z + d_2 \mathbbm{1})  \notag , 
\end{align}
and likewise for $J_\uparrow^y(\k)$. 
 \\

\section{Evaluation of the drag diagrams in the continuum model}
\label{appdragsec}

Let us start by considering the first ``direct'' diagram in Fig.\ \ref{dragdiags} with majority band indices $\alpha = 1, \beta = 2$. Its contribution to the Matsubara correlator $ - \braket{\hat{J}_\downarrow^x \hat{J}_\uparrow^y}(i\Omega)$, to be denoted by $\mathcal{P}_1(i\Omega)$, reads
\begin{widetext}
\begin{align}
&\mathcal{P}_1(i\Omega) = -g^2n_\downarrow A_0 \int_{k,q} \int \frac{d\tilde{\omega}}{2\pi} J_{\uparrow,12}^y(\k) W^{22}(\k-\q,\q) W^{21}(\k,-\q) J^x_{\downarrow}(\q) \\ & \frac{1}{i(\Omega + \omega_q) - \epsilon_\downarrow(\q)} \frac{1}{i\omega_q - \epsilon_\downarrow(\q)} \frac{1}{i(\omega_q + \tilde{\omega}) + 0^+ } \frac{1}{i(\Omega + \omega_k) - \eone(\k)} \frac{1}{i\omega_k - \etwo(\k)} \frac{1}{i(\omega_k + \tilde{\omega}) -\etwo(\k - \q)} \notag  \ ,  
\end{align}
where $0^+$ in the third impurity propagator ensures the correspondence to filled states. Evaluating the frequency integrals we find: 
\begin{align}
&\mathcal{P}_1(i\Omega) = -g^2n_\downarrow A_0 \int \frac{d\k}{(2\pi)^2} \frac{d\q}{(2\pi)^2} J_{\uparrow,12}^y(\k) W^{22}(\k-\q,\q) W^{21}(\k,-\q) J^x_{\downarrow}(\q) \\ & \frac{1}{\eone(\k) - \etwo(\k-\q) - \epsilon_\downarrow(\q)} \frac{1}{-i\Omega + \eone(\k) - \etwo(\k - \q) - \epsilon_\downarrow(\q)}\frac{1}{-i\Omega + \eone(\k) - \etwo(\k)} \notag \ . 
\end{align}
Upon analytical continuation, $i\Omega \rightarrow \omega$, only the $\mathcal{O}(\omega)$ part contributes to the static drag as in the non-interacting case. With Eq.\ \eqref{dragtrans1}, we get: 
\begin{align} 
\label{sigma1app}
\sigma_{\downarrow\uparrow,1} = 
 i g^2 n_\downarrow \int \frac{d\k}{(2\pi)^2} \frac{d\q}{(2\pi)^2}  J^y_{\uparrow,12}(\k) W^{22}(\k - \q, \q) W^{21}(\k,-\q)  \frac{q_x}{M} \frac{1}{\left(\eone(\k) - \etwo(\k)\right)^2} d(\k,\q), 
 \end{align}
 \end{widetext}
with $d(\k,\q)$ as defined in Eq.\ \eqref{cddef}. The remaining three contributions to $\sigma_{\downarrow\uparrow}$ have the following structure: The direct diagram with majority indices $\alpha = 2, \beta = 1$ leads to Eq.\ \eqref{sigma1app} with $A \equiv  J^y_{\uparrow,12}(\k) W^{22}(\k - \q, \q) W^{21}(\k,-\q) $ replaced by $B \equiv - J^y_{\uparrow, 21}(\k) W^{12}(\k - \q,\q) W^{22}(\k,-\q)$; using elementary properties of unitary matrices, one can show that $B = -\overline{A}$ (with $\overline{A}$ the complex conjugate of $A$), thus yielding the part $\propto d(\k,\q)$ of Eq.\ \eqref{maineqdragcont} in the main text. The remaining ``crossed'' diagram of Fig.\ \ref{dragdiags} likewise generates the part $\propto c(\k,\q)$. 
\\

\section{Jump of the Hall drag in the continuum model}
\label{contjumpapp} 
To derive the jump from Eq.\ \eqref{maineqdragcont}, we need to project on the part of the $\k$-integrand corresponding to the Dirac fermions, which becomes singular at $\k = 0$ as $m \rightarrow 0$. This can be done by setting $\k = 0$ in all regular parts. The last factor in the integrand becomes
\begin{align}
\label{cdsimpl}
&d(\k,\q) + c(\k,\q) \rightarrow  \\ \notag 
&\left( \frac{1}{(\epsilon_\downarrow(\q) + \etwo(\q))^2} -  \frac{1}{(\epsilon_\downarrow(\q) - \eone(\q))^2} \right) \  .
\end{align}
In the part involving interaction matrices $W$, it is useful to rewrite 
\begin{align}
&W^{22}(\k - \q, \q) W^{21}(\k,-\q) \overset{\eqref{Hint}}{=} \\ & \notag U^\dagger_{\uparrow,2n}(\k) U_{\uparrow,n2}(\k-\q) U^\dagger_{\uparrow,2 m} (\k - \q) U_{\uparrow,m1}(\k) \rightarrow \notag  \\& \notag U^\dagger_{\uparrow,2n}(\k) U_{\uparrow,n2}(-\q) U^\dagger_{\uparrow,2 m} (- \q) U_{\uparrow,m1}(\k) =  \\ & \notag \left(U^\dagger_{\uparrow}(\k) V(\q) U_\uparrow(\k) \right)_{21} , \\&  \notag V(\q)_{nm} \equiv  U_{\uparrow,n2}(-\q) U^\dagger_{\uparrow,2 m} (- \q) \ , 
\end{align}
where $n,m$ are sublattice indices, and in the second step we have only kept the singular $\k$ dependence. $V(\q)$ is a hermitian matrix, and so can be expanded as a linear combination of the unit and Pauli matrices with real coefficients. Then it is easy to show that only the contribution $\propto \sigma_x$ survives the integration in Eq.\ \eqref{maineqdragcont}, while the rest either does not contribute to the required imaginary part or is antisymmetric in $k_x$. Therefore, we can write 
\begin{align}
\label{UJUsimpl}
& \left(U^\dagger_{\uparrow}(\k) V(\q) U_\uparrow(\k) \right)_{21} \hat{=}  \\  \notag & \left(U^\dagger_{\uparrow}(\k ) \sigma_x U_\uparrow(\k) \right)_{21} \text{Re} \left[V(\q)_{12}\right] = \\& \notag \left(U^\dagger_{\uparrow}(\k)  J^{x,0}_{\uparrow,\text{Dirac}}U_\uparrow(\k) \right)_{21} \frac{-q_x}{2q\sqrt{1 + (d_1 q)^2} } \ ,  \end{align}
where in the last step we identified the current operator of the Dirac fermions in the sublattice basis, $\sigma_x = J^{x,0}_{\uparrow,\text{Dirac}}$ (cf.\ Eq.\ \eqref{contham}), and wrote out $V(\q)$ by inserting matrix elements of $U_\uparrow(-\q)$ from App.\ \ref{BasisrotSec}. Inserting Eqs.\ \eqref{cdsimpl}, \eqref{UJUsimpl} into Eq.\ \eqref{maineqdragcont}, we can write the sign-changing Dirac part of the Hall drag as shown in Eq.\ \eqref{sigmaDiracfirst} of the main text.

\section{Impurity Hall drag and jump in the Haldane model}
\label{Haldaneapp}

In the Haldane model, the on-site interaction is defined by (cf.\ Eq.\ \eqref{Hint})
\begin{widetext}
 \begin{align}\notag
&H_\text{int} = \frac{g}{A_0}\sum_{\ell = A,B}\sum_{\k,\p,\q}  c^\dagger_{\uparrow, \ell} (\k +\q) c_{\uparrow,\ell} (\k) c^\dagger_{\downarrow, \ell}(\p-\q) c_{\downarrow, \ell}(\p) =  \frac{g}{A_0} \sum_{\k,\p,\q} c^\dagger_{\uparrow, \alpha} (\k +\q) c_{\uparrow,\beta} (\k) c^\dagger_{\downarrow,1}(\p-\q) c_{\downarrow,1}(\p) W_{\alpha\beta}(\k,\p,\q)\ ,  \\ &
 W_{\alpha \beta} (\k, \p, \q) = \sum_{\ell = A,B} \overline{U}_{\uparrow, \ell \alpha}(\k + \q) U_{\uparrow, \ell \beta}(\k) 
\overline{U}_{\downarrow, \ell 1}(\p - \q) U_{\downarrow, \ell 1}(\p) \ . 
\label{HintHaldane}
\end{align}
In Eq.\ \eqref{HintHaldane}, we have restricted the impurity to the lower band, which is legitimate for weak interactions.

With this interaction, the derivation of the Hall drag proceeds analogously to the  continuum model, App.\ \ref{appdragsec}, and results in 
\begin{align}
\label{haldanedragformula}
&\sigma_{\downarrow\uparrow} = \\ \notag &-2 g^2 n_\downarrow \int \frac{d\k}{(2\pi)^2} \frac{d\q}{(2\pi)^2} \  \text{Im} \left\{ J^y_{\uparrow,12}(\k) W^{22}(\k - \q, \q, \q) W^{21} (\k, \0, -\q)\right\}  J^x_{\downarrow,11}(\q) \frac{1}{\left(\eone(\k) - \etwo(\k)\right)^2}  \left(d(\k,\q) + c(\k,\q) \right) , 
\end{align}
with $J_{\downarrow,11}^x(q)$ the impurity current operator in the band basis (taking into account lower band contributions only), and $c,d$ as in Eq.\ \eqref{cddef}, only replacing the single band energy of the continuum model $\epsilon_{\downarrow}(\q)$ by the lower band energy $\epsilon_{\downarrow,1}(\q)$. 

From Eq.\ \eqref{haldanedragformula} one can readily derive the additional symmetry $\sigmad(\phi) = - \sigmad(\pi -\phi)$ mentioned in the main text. In the majority Hamiltonian $H_\uparrow(\k)$, $h_0(\k; \phi) = - h_0(\k; \pi - \phi)$, while the other coefficients are invariant under such reflection. As a result, one finds $c(\k,\q; \phi) = - d(\k, \q; \pi - \phi)$. All other elements of Eq.\ \eqref{haldanedragformula} do not change, which shows the property as claimed.

To evaluate the jump of the Hall drag $\Delta\sigmad$ in the Haldane model in analogy with Sec.\ \ref{jumpsec}, let us focus on the transition line, $ \Delta_c  =  6\sqrt{3} t^\prime \sin(\phi)$, where the gap closes at the Dirac point $\k_A = (0,4\pi/3\sqrt{3})^T$. Since $\sigmad$ is symmetric in $\Delta$, for a given value of $\phi$ the value of $\Delta\sigmad$ at $- \Delta_c$ is the same. To extract the singular Dirac contribution at $\k_A$, we let $\k \rightarrow \k_A$ in all regular parts of Eq.\  \eqref{haldanedragformula}. In this limit, \begin{align}  \label{jneed}
J_\uparrow^x(\k) \rightarrow U_\uparrow^\dagger(\k) \frac{3}{2} \sigma_y U_\uparrow(\k) \equiv J_{\uparrow, \text{Dirac}}^x(\k)  \end{align}
This current can be extracted from the interaction part of Eq.\ \eqref{haldanedragformula} as in Sec.\ \ref{jumpsec}, which allows to write the Dirac part  of the Hall drag as
\begin{align}
\label{sigmadirachald}
&\sigma_{\downarrow\uparrow,\text{Dirac}} =  \frac{g^2 n_\downarrow}{(2\pi)^4}  \int \frac{d\k}{\pi}  \ \text{Im} \left\{ J_{\uparrow, \text{Dirac},12}^y(\k) J_{\uparrow, \text{Dirac}, 21}^x(\k) \right\} \cdot \frac{1}{\left(\eone(\k) - \etwo(\k)\right)^2} \cdot f(t^\prime, \phi) \ , \\ \notag & f(t^\prime, \phi)  \equiv  -\frac{4\pi}{3}\int  d\q  J_{\downarrow}^x(\q) \left(\frac{1}{\left(\eone(\k_A)- \etwo(\k_A-\q)-\Eone(\q)\right)^2} -  
\frac{1}{\left(\eone(\k_A-\q)-\etwo(\k_A) -\Eone(\q)\right)^2} \right) \cdot \\ \notag  &\qquad \qquad \qquad \text{Im} \left\{ U_{\uparrow, A2}(\k_A - \q) U^\dagger_{\downarrow,1A} (\0) U_{\downarrow,A1} (\q) U^\dagger_{\uparrow,2B}(\k_A - \q) U_{\downarrow,B1} (\0) U^\dagger_{\downarrow,1B} (\q) \right\} \quad 
  \end{align}
Again, the $\k$ and $\q$ integrals have factorized, and the $\k$ integral gives $\pm 1/2$. This yields a value of the jump 
as in Eq.\ \eqref{haldanejumpmain} of the main text. The remaining $\q$ integral has to be evaluated numerically. 
\end{widetext}

\section{Antisymmetry of the hall drag as function of $g$ in the Haldane model with $\phi = \pm \pi/2, \Delta = 0$. }
\label{antapp}

Here we show that the Hall drag $\sigmad$ in the Haldane model, with parameters $\phi = \pm \pi/2, \Delta = 0$, is antisymmetric in the impurity-majority coupling $g$ to all orders. We work in the diagonal band frame,  and perform a particle-hole transformation which also exchanges the band indices: 
\begin{align}
\label{phflip}
  b_{\tilde{\alpha}}(\k) \equiv c^\dagger_{\uparrow,\alpha}(-\k) , \quad 
{b}^\dagger_{\tilde{\alpha}}(\k) \equiv {c}_{\uparrow,\alpha}(-\k)  , \quad {\tilde{1}} \equiv 2,  \quad \tilde{2} \equiv 1\ . 
\end{align}
Due to particle-hole symmetry for $\phi = \pm \pi/2$, the form of the non-interacting majority Hamiltonian is invariant under this transformation (up to a constant): 

\begin{align}
\label{Huptrafo}
H_\uparrow &= \sum_\k c^\dagger_{\uparrow,\alpha}(\k) \epsilon_{\alpha}(\k) c_{\uparrow,\alpha}(\k)  \\ \notag &= \sum_\k  b^\dagger_{\tilde{\alpha}}(-\k) 
\left[- \epsilon_{\alpha}(\k)\right] b_{\tilde{\alpha}}(-\k)  + \text{const.} \\ \notag &
 = \sum_\k b^\dagger_{\alpha}(\k)  
 \epsilon_{\alpha}(\k) b_{\alpha}(\k)  + \text{const.} \ , 
 \end{align}
 where $\epsilon_{\alpha}(\k) = - \epsilon_{\tilde{\alpha}}(-\k)$ was used.
However, the interaction term acquires a minus sign under the variable transformation \eqref{phflip}:
 \begin{widetext} 
 \begin{align}
 \label{Hinttrafro}
 H_\text{int}   &=   \frac{g}{A_0} \sum_{\k,\p,\q} c^\dagger_{\uparrow, \alpha} (\k +\q) c_{\uparrow,\beta} (\k) c^\dagger_{\downarrow,1}(\p-\q) c_{\downarrow,1}(\p) W_{\alpha\beta}(\k,\p,\q)  \\ & \notag= -  \frac{g}{A_0} \sum_{\k,\p,\q} b^\dagger_{\tilde\beta}(-\k) {b}_{\tilde \alpha}(-\k-\q)  c^\dagger_{\downarrow,1}(\p-\q) {c}_{\downarrow,1}(\p) W_{\alpha \beta} (\k, \p, \q)  + \text{const.}  \\ \notag &  = 
   - \frac{g}{A_0} \sum_{\k,\p,\q}{b}^\dagger_{\alpha}(\k+\q) {b}_{\beta}(\k)  c^\dagger_{\downarrow,1}(\p-\q) {c}_{\downarrow,1}(\p)  W_{\tilde\beta \tilde\alpha} (-\k - \q, \p, \q) + \text{const.}   \\ &  =  \notag - \frac{g}{A_0} \sum_{\k,\p,\q} b^\dagger_\alpha (\k +\q) b_{\beta} (\k) c^\dagger_{\downarrow,1}(\p-\q) c_{\downarrow,1}(\p) W_{\alpha\beta}(\k,\p,\q) + \text{const.} \ .  \end{align}
 \end{widetext}
The unimportant additional terms are constant in the majority sector. In the last step, we used $W_{\tilde\beta\tilde\alpha}(-\k - \q, \p, \q) = W_{\alpha\beta}(\k,\p,\q)$. This can be easily shown by inserting the matrix elements from Eqs.\ \eqref{Uall}, \eqref{Haldaneham}, but requires $h_3(\k) = - h_3(-\k)$, which is only fulfilled for $\Delta = 0$ (and is violated in the continuum model). 
  
Last, the required majority current operator transforms as 
\begin{widetext}
\begin{align}
&J_{\uparrow, \alpha \beta}^y(\k)= \sum_\k c^\dagger_{\uparrow,\alpha}(\k) U^\dagger_{\uparrow,\alpha n}(\k) \left[J_{y, \uparrow}^0(\k)\right]_{nm}  U_{\uparrow, m \beta}(\k) c_{\uparrow,\beta}(\k) , \quad \left[J_{y, \uparrow}^0(\k)\right]_{nm} = \left[ \partial_{k_y} H_\uparrow(\k) \right]_{nm} \ , \\ \notag 
&J_{\uparrow, \alpha \beta}^y(\k) = - \sum_{\k} {b}^\dagger_{\tilde\beta}(-\k) U^\dagger_{\uparrow,\alpha n}(\k)  \left[J_{y, \uparrow}^0(\k)\right]_{nm}  U_{\uparrow, m \beta}(\k) b_{\tilde\alpha}(-\k) + \text{const.} \\  &\quad \quad \quad \ \  =  - \sum_{\k} {b}^\dagger_{\alpha}(\k) U^T_{\uparrow,\tilde\alpha m}(-\k)  \left[J_{y, \uparrow}^0(-\k)\right]^T_{mn}  \overline{U}_{\uparrow, n \tilde \beta}(-\k) b_{\beta}(\k) + \text{const.} \notag \ . 
\end{align}
\end{widetext}
Again, inserting matrix elements one can show that 
\begin{align*}
&U^T_{\uparrow,\tilde\alpha m}(-\k)  \left[J_{y, \uparrow}^0(-\k)\right]^T_{mn}  \overline{U}_{\uparrow, n \tilde \beta}(-\k) =  \\ &U_{\uparrow,\alpha m}^\dagger(\k) \left[J_{y, \uparrow}^0(\k)\right]_{mn}  U_{\uparrow, n \beta}(\k), \end{align*}  and the majority current changes sign. In conclusion, for $\phi = \pm \pi/2, \Delta = 0$ this proves the antisymmetry \begin{align}
\sigmad(g) = - \sigmad(-g) \ , 
\end{align}
as claimed in the main text.

\section{$\sigmad$ from circular dichroism: Technical details}
\label{dichroapp}

The Feynman rules for the perturbation $H_{\uparrow,\pm}(t)$ of Eq.\ \eqref{Hpmup} in the energy-momentum domain are easily derived from Wick's theorem. They read: 
 
 \begin{itemize} 
\item Each current vertex comes with a factor $E/\omega$. 
\item If an incoming (outgoing) electrical field line couples to a $J_x$-vertex, there is an extra factor $-i$ $(i)$ for both $\Gamma_{\pm}(\omega)$. 
\item If an electrical field line (incoming or outgoing)  couples to a $J_y$-vertex, this gives a factor $\mp$ 1 for $\Gamma_\pm(\omega)$. 
\end{itemize} 
Application of these rules directly leads to Eq.\ \eqref{DeltaGammasome} in the non-interacting case. For the integrated impurity depletion rate, let us consider for instance the contribution of the two diagrams of Fig.\ \ref{imp_rate}(c), \ref{imp_rate}(d), to be denoted $D$. It reads

\begin{widetext}
\begin{align}
&D = - n_\downarrow g^2 E^2A_0   \int_0^\infty d\omega \int \frac{d\k}{(2\pi)^2}  \frac{d\q}{(2\pi)^2} \ \text{Im}\  \bigg\{\int \frac{d\omega_k}{2\pi} \int \frac{d\omega_q}{2\pi} \left(- 2iJ^y_{\uparrow,21}(\k) J^x_\downarrow(\q)W^2 +2iJ^y_{\uparrow,12}(\k)J^x_\downarrow(\q) \overline{W}^2 \right) \frac{1}{\omega^2}  \\ &  \frac{1}{\omega_q - \edown(\q) + i0^+} \frac{1}{\omega_q - \omega - \edown(\q)+ i0^+} \frac{1}{\omega_k - \eone(\k) - i0^+} \frac{1}{\omega + \omega_k -\etwo(\k) +i0^+} \frac{1}{\omega + \omega_k - \omega_q -\etwo(\k-\q) + i0^+} \bigg\}  \notag \ . 
\end{align}
Here $W$ is shorthand for the proper interaction matrices (cf.\ Eq.\ \eqref{maineqdragcont}). The third propagator is advanced (it corresponds to a majority hole) and has a $-i0^+$ term in the denominator, the other propagators are retarded. Performing the $\omega_k, \omega_q$ integrals yields 
\begin{align}
&D = -n_\downarrow g^2E^2A_0\ \int \frac{d\k}{(2\pi)^2}  \frac{d\q}{(2\pi)^2}   \int_{>0}d\omega  \ \text{Im}\  \bigg\{ \left(- 2iJ^y_{\uparrow,21}(\k) J^x_\downarrow(\q)W^2 +2iJ^y_{\uparrow,12}(\k)J^x_\downarrow(\q) \overline{W}^2 \right) \frac{1}{\omega^2}  \\ &  \frac{1}{\omega + \eone(\k) - \etwo(\k-\q) - \edown(\q) + i0^+} \frac{1}{\eone(\k) - \edown(\q) - \etwo(\k-\q)+ i0^+} \frac{1}{\omega + \eone(\k) -\etwo(\k) + i0^+} \bigg\} \notag \ . 
\end{align}
The expression involving the currents is real, and the imaginary part comes from the propagators only. They yield a sum of two delta-functions, since the propagator in the middle is real. Computing the $\omega$-integral, after some trivial algebra one then finds

\begin{align}
&D = 2\pi E^2 A_0 \cdot - 2g^2 n_\downarrow  \int  \frac{d\k}{(2\pi)^2} \frac{d\q}{(2\pi)^2}  \ \text{Im}\left
\{J^y_{\uparrow,12}(\k) J_\downarrow^x(\q) W^2 \right\} \frac{2 \eone(\k) - \etwo(\k) -\etwo(\k-\q) - \edown(\q)}{(\etwo(\k) - \eone(\k))^2 (\eone(\k) - \etwo(\k-\q) - \edown(\q))^3}\  , 
\end{align}
which is precisely $2\pi E^2 A_0$ times the $\sigmad$-contribution of the ``direct'' diagram, cf.\ \eqref{maineqdragcont}. Evaluation of the other non-vanishing drag diagrams (crossed diagram and diagrams with $J_\downarrow^y, J_\uparrow^x$ interchanged) proceeds in the same manner.

Since diagrams where both external field lines couple to the impurity vanish when forming $\Delta \Gamma_{\downarrow}$, the only remaining non-zero diagrams are those of  Fig.\ \ref{imp_rate}(a),  \ref{imp_rate}(b) plus those with inverted directions of the external field lines. After some straightforward simplifications, one finds a total contribution 

\begin{align}
\label{extraEq}
& \frac{n_\downarrow g^2}{(2\pi)^4}4 E^2A_0 \  \int_0^\infty \frac{d\omega}{\omega^2}  \int d\k d\q \  \text{Im} \left[ J_{\uparrow,21}^x(\k- \q) J^y_{\uparrow,12}(\k) W_{11}(\k,-\q,-\q) W_{22}(\k-\q,\0,\q) \right]  \\ & \notag \text{Im} \bigg\{ (-1) \frac{1}{-\omega + \etwo(\k-\q) - \eone(\k - \q) - i0^+} \frac{1}{\omega + \eone(\k) - \etwo(\k) + i0^+}  \\& \notag \cdot \left(\frac{1}{\omega - \etwo(\k) + \eone(\k - \q) - \edown(\q) + i0^+} +  \frac{1}{\omega + \eone(\k) - \etwo(\k - \q) - \edown(\q) + i0^+}  \right) \bigg\} \ . 
\end{align}
\end{widetext} 
It is readily seen that this expression is invariant under $\eone \leftrightarrow - \etwo$, which implies the particle-hole symmetry claimed in the main text. We have also checked this symmetry explicitly for the Haldane model by numerically implementing  Eq.\ \eqref{extraEq}.


\bibliographystyle{apsrev4-1}

\bibliography{../top_polaron}

\end{document}